\documentstyle[aps,epsfig,amstex,twocolumn,subfigure]{revtex}
\newcommand{\ve}[1]{{\mathbf{#1}}}
\newcommand \bea{\begin{eqnarray}}
\newcommand \eea{\end{eqnarray}}
\newcommand \ga{\raisebox{-.5ex}{$\stackrel{>}{\sim}$}}
\newcommand \la{\raisebox{-.5ex}{$\stackrel{<}{\sim}$}}
\begin{document}
\title{Cooper pairing and single particle properties of trapped Fermi gases}

\author{G.\ M.\ Bruun and H.\ Heiselberg}
\address{Nordita, Blegdamsvej 17, 2100 Copenhagen, Denmark}
\maketitle

\begin{abstract}
We calculate the elementary excitations and pairing of a trapped
atomic Fermi gas in the superfluid phase. The level spectra and pairing
gaps undergo several transitions
as the strength of the interactions between and the number of atoms are
varied. For weak interactions, the Cooper pairs are formed between
particles residing in the same harmonic oscillator shell. In this
regime, the nature of the paired state is shown to depend critically
on the position of the chemical potential relative to the harmonic
oscillator shells and on the size of the mean field. For stronger
interactions, we find a region where pairing occur between
time-reversed harmonic oscillator states in different shells also.
\end{abstract}

\section{Introduction}
The trapping and cooling of trapped fermionic atomic gases is becoming
a very active area of experimental research. Presently, temperatures
as low as $\sim 0.2T_F$ have been obtained for $^{40}$K~\cite{DeMarco} and
$^6$Li~\cite{Truscott,Schreck,O'Hara} with $T_F$ denoting the Fermi
temperature. Part of the motivation for this impressive experimental
progress is to observe the theoretically predicted transition to a
superfluid state below a critical temperature $T_c$. With a large and
attractive low energy effective interaction between two hyperfine
states trapped, $T_c$ for such a transition should be experimentally
obtainable~\cite{StoofBCS}. 

The purpose of the present paper is
examine the elementary excitations of the ground state for an atomic
superfluid gas in a spherical harmonic trap. 
Varying the interaction strength and the number of particles in the trap
a number of interesting transitions in the level spectra and pairing gaps
occur. 
We calculate these transitions by solving the Bogoliubov-deGennes
equations both numerically and analytically using  a degenerate shell pairing approximation  for weak 
interactions and semiclassical methods for many particles.
An understanding of
such elementary excitations is important if one wants to predict
the response of the gas to various single particle probes. Moreover,
these elementary excitations are also the building blocks, if one
wants to calculate collective properties of the superfluid gas. 

The paper is organized as follows: In sec.~\ref{Basicstuff}, we set up
the basic formalism used in the rest of the paper. We then in
sec.~\ref{intrashellsection} discuss the weak pairing regime, where
Cooper pairs are formed only between particles within the same
harmonic oscillator shell. The qualitatively different limit of
strong pairing is discussed in sec.~\ref{Strongpairing}, where Cooper
pairs are formed between particles residing in different shells. We
finally summarize our results in sec.~\ref{conclusion}.

\section{Basic equations}\label{Basicstuff}
We consider a dilute gas of $N$ fermionic atoms of mass $m$ in two hyperfine
states $|\sigma=\uparrow,\downarrow\rangle$ trapped by a spherically
symmetric harmonic oscillator (h.o.) potential at zero temperature.
The numbers $N_\sigma$ of atoms trapped in each hyperfine state are
assumed to be equal such that $N=2N_\sigma$ as this is the optimum situation for Cooper
pairing.  We assume that the atoms interact through an attractive
s-wave scattering length, $a<0$. The Hamiltonian is then given by
\bea\label{Hamiltonian}
  H &=& \sum_{i=1}^{N} \left( \frac{{\bf p}_i^2}{2m} +
    \frac{1}{2} m\omega^2 r_i^2 \right)
  + 4\pi \frac{\hbar^2a}{m} \sum_{i<j} 
     \delta({\bf r}_i-{\bf r}_{j}) \,.
\eea
It should be noted that to reproduce the correct 2-body scattering properties one needs to use 
a pseudo-potential $\delta({\bf r})\partial_rr$ instead of a simple delta-function for the 
interaction term in Eq.(\ref{Hamiltonian}).  This is important when the pairing correlations are to be calculated~\cite{BruunBCS}.
 We will return to this point in sec.\ \ref{intrashellsection}. In a non-interacting system the Fermi energy is
\bea
   E_F=(n_F+3/2)\hbar\omega \,
  \simeq \, \left(3N\right)^{1/3} \hbar\omega   \,, \label{EF}
\eea
where $n_F$ is the h.o. quantum number at the Fermi surface.
For a large number of particles $n_F=(3N)^{1/3}$.

Interactions result in a mean field potential which is
for a large number of particles given by the Thomas-Fermi 
approximation (TF)
\bea
    U(r) &=& 2\pi\frac{\hbar^2a}{m} \rho(r) \,, \label{U}
\eea
where the particle density is
\bea
     \rho(r) = k_F^3(r)/3\pi^2 \,
     \simeq \rho_0 \left(1-r^2/R_{TF}^2-U(r)/E_F\right)^{3/2} \, \label{rho} 
\eea
inside the cloud $r\le R_{TF}=a_{osc}\sqrt{2n_F+3}$, where 
$a_{osc}=\sqrt{\hbar/m\omega}$ is the oscillator length, and
$\rho_0=(2n_F)^{3/2}/3\pi^2 a_{osc}^3$ is the central density~\cite{Butts}. 
$k_F(r)$ is the TF Fermi wave-number
and the Fermi energy is
\bea
   E_F = \hbar^2k_F^2(r)/2m +(1/2)m\omega^2r^2+U(r). 
\eea
In dilute systems the interactions contribute a mean field weak compared to the
confining potential, i.e., $U(r)$ can be neglected in Eq.~(\ref{rho}).

 The system is {\it dilute} when the Fermi energy is large compared to
the mean field energy, $E_F\gg U$, or equivalently $k_F|a|\simeq
\rho^{1/3}|a|\simeq n_F^{1/2}|a|/a_{osc}\ll 1$.  {\it Dense}
Fermi systems, $k_F|a|\ga 1$, are studied in Refs. \cite{HH}.
In {\it very dilute}
systems also the h.o.  energy exceeds the mean field
potentials, $\hbar\omega\gg U$, or equivalently
$n_F^{3/2}|a|/a_{osc}\ll 1$. 

The h.o. levels are highly degenerate with states having angular momenta
$l=n, n-2\ldots ,0$ ($n$ even) or $l=n, n-2 \ldots ,1$ ($n$ odd)
due to the   $U(3)$ symmetry of the 3D spherically symmetric h.o. potential.
However, interactions split this degeneracy.
The splitting 
can be calculated perturbatively in the very dilute limit.
From the radial h.o. wave-function ${\cal R}_{n_Fl}(r)$ for the state
with angular momentum $l$ and $(n_F-l)/2$ radial nodes in the h.o. shell
$n_F$, the single particle energies with respect to the
Fermi energy are within the WKB approximation~\cite{Heiselberg}
\bea \label{Hartree}
  \xi_{n_F,l} &=& \int U(r) 
 |{\cal R}_{nl}(r)|^2 r^2dr \nonumber\\
  &=& \frac{4\sqrt{2}}{3\pi}\frac{a}{a_{osc}} n_F^{3/2} \hbar\omega
\left[\frac{4}{3\pi}-\frac{l(l+1)}{4\pi n_F^2}\right]
 \,.
\eea
The mean field energies are proportional to the coupling and $n_F^{3/2}$
and are split like rotational bands with a prefactor that
is smaller than the average mean field
energy by a factor $\sim3/16$. The relative small splitting
reflects the fact that the mean field potential is almost
quadratic in $r$  and the anharmonic terms therefore small.

We plot in figure 1 the
different regions of pairing for the trapped gas as a function of the
number of particles and the coupling strength.
\begin{figure}
\begin{center}
\psfig{file=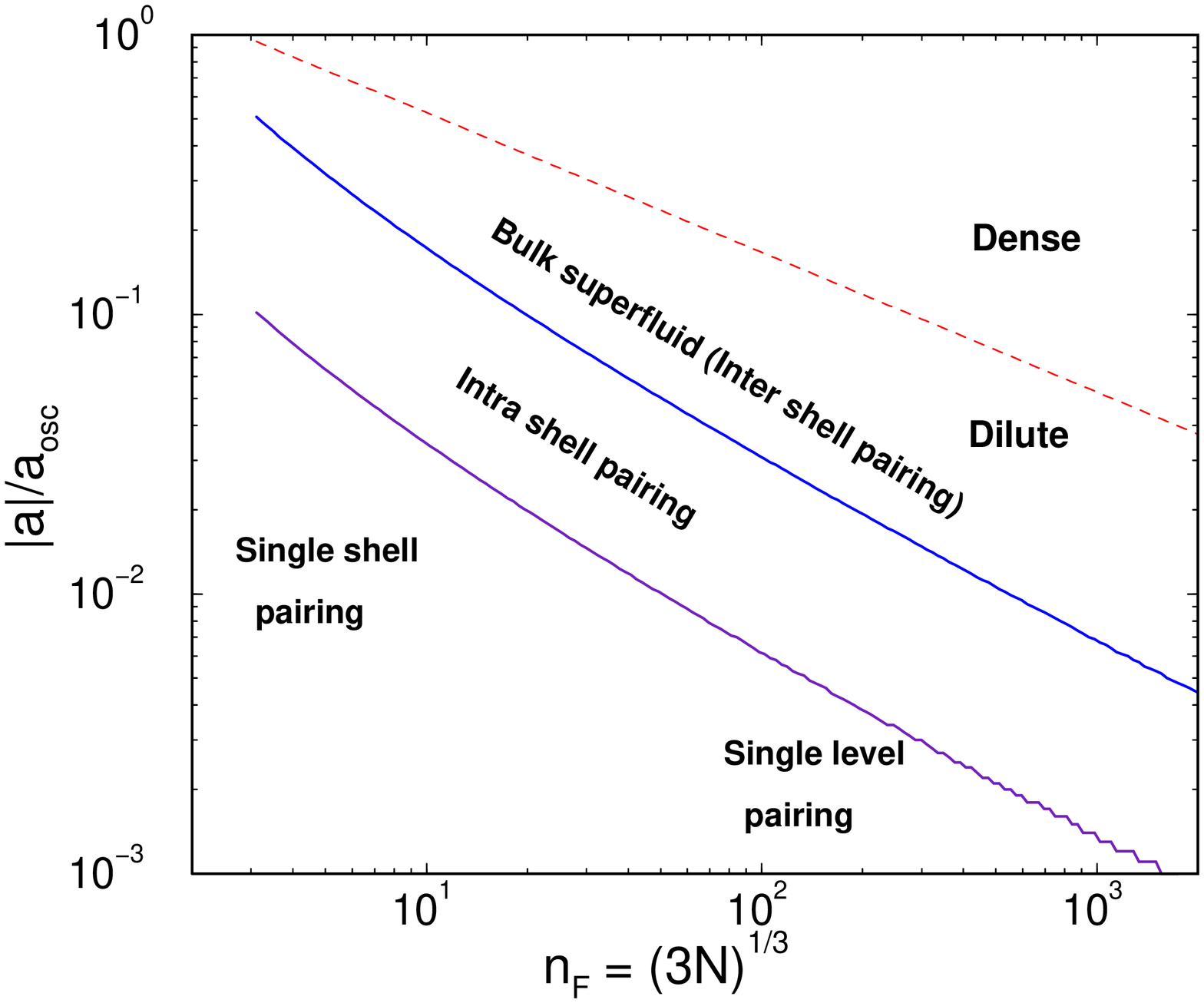,height=8.0cm,angle=-90}
\vspace{.2cm}
\begin{caption}{The different level structures and pairing gaps in traps are
displayed vs.  the two dimensionless parameters: The energy quantum
number $n_F$ of the highest shell occupied and the interaction
strength $|a|/a_{osc}$.  The dotted line separates the dense and dilute
trapped gases.  Above the blue full line $\Delta=\hbar\omega$, the system
has coherence length smaller than the system size and pairing occurs
between particles in different harmonic oscillator shells. Below, a
region exists where there is only pairing between particles in the
same shell but many h.o. shells contribute to the total pairing field.
Below the purple line $G\ln(n_F)\simeq0.1\hbar\omega$, there is only
pairing in the shell at the chemical potential leading to the gap
$\Delta\simeq G$ for small systems whereas pairing takes place in a single $l$-level only
for large systems. It should be noted that the transitions between the 
different regions are smooth.}
\end{caption}
\end{center}
\label{phase}
\end{figure}
 We shall below calculate pairing energies by solving the
Bogoliubov-de Gennes (BdG) equations numerically and also
analytically in certain limits exploiting known properties of the known h.o.
wave-functions and single particle levels. The BdG equations
read~\cite{deGennes}:
\begin{eqnarray}\label{BdGeqn}
 E_\eta u_\eta({\mathbf{r}}) &=& [{\cal H}_0 + U({\mathbf{r}})]
 u_\eta({\mathbf{r}}) +\Delta({\mathbf{r}}) v_\eta({\mathbf{r}})\nonumber \\
E_\eta v_\eta({\mathbf{r}}) &=& -[{\cal H}_0 + U({\mathbf{r}})]v_\eta({\mathbf{r}}) +
\Delta({\mathbf{r}}) u_\eta({\mathbf{r}})\label{BdG}.
\end{eqnarray}
Here, ${\mathcal{H}}_0= -(\hbar^2/2m)\nabla^2+U_0({\mathbf{r}})-\mu_{\rm F}$,
 $U({\mathbf{r}})\equiv g\langle\hat{\psi}_{\sigma}^{\dagger}({\mathbf{r}})
\hat{\psi}_{\sigma}({\mathbf{r}})\rangle=g\rho_\sigma({\mathbf{r}})$ 
is the Hartree potential, and 
 $\hat{\psi}_{\sigma} ({\mathbf{r}})$ is the field operator that annihilates an atom 
in spin state  $\sigma$ at position   $\ve{r}$.  The coupling constant  is 
 $g=4\pi a\hbar^2/m$ with $a<0$ for an attractive interaction leading to pairing, 
and $\mu_F$ denotes the chemical potential. The pairing field is defined as
 $\Delta({\mathbf{R}})\equiv -g\lim_{r\rightarrow 0}\partial_r[r
\langle\hat{\psi}_\uparrow({\mathbf{R}}+{\mathbf{r}}/2)
\hat{\psi}_\downarrow({\mathbf{R}}-{\mathbf{r}}/2)\rangle]$ and can be taken to
be real~\cite{BruunBCS}.
The quasi particles with energies $E_\eta$ are described
 by the Bogoliubov wave  functions $u_\eta({\mathbf{r}})$ and $v_\eta({\mathbf{r}})$. 

For spherical symmetry, the Bogoliubov wave functions from Eq.(\ref{BdGeqn}) have 
definite  angular momentum quantum numbers $lm$ and can be written in the form 
$u_{\eta lm}({\mathbf{r}})= r^{-1}u_{\eta l}(r)Y_{lm}(\theta,\phi)$ and 
$v_{\eta lm}({\mathbf{r}}) = r^{-1}v_{\eta l}(r)Y_{lm}(\theta,\phi)$ with an 
energy energy $E_{\eta l}$. To obtain analytical results for the pairing energy in some 
limits, it is convenient to expand the Bogoliubov wavefunctions in the 
complete set of unperturbed spherical Harmonic oscillator functions  
by writing
$u_{\eta lm}({\mathbf{r}})=\sum_{n}u_{\eta nl}\phi^0_{nlm}({\mathbf{r}})$
and $v_{\eta lm}({\mathbf{r}})=\sum_{n}v_{\eta nl}\phi^0_{nlm}({\mathbf{r}})$ with 
$\phi_{nlm}({\mathbf{r}})={\mathcal{R}}_{n l}(r)Y_{lm}(\theta,\phi)$. We have 
${\mathcal{H}}_0\phi^0_{nlm}({\mathbf{r}})=\xi_n\phi^0_{nlm}({\mathbf{r}})$ with 
 $\xi_n=(n+3/2)\hbar\omega-\mu_F$ [$l=n, n-2\ldots ,0$ ($n$ even) or $l=n, n-2 \ldots ,1$ ($n$ odd)]. 
Using this expansion,  the BdG equations become for a given $lm$:
\begin{eqnarray}
\label{BdGeqnexpand}
E_{\eta l}u_{\eta nl}&=&
\sum_{n'}[\langle nl|{\mathcal{H}}|n'l\rangle u_{\eta n'l}+
\langle nl|\Delta(r)|n'l\rangle v_{\eta n'l}]\nonumber \\
E_{\eta l}v_{\eta nl}&=&
\sum_{n'}[\langle nl|\Delta(r)|n'l\rangle u_{\eta n'l}
-\langle nl|{\mathcal{H}}|n'l\rangle v_{\eta n'l}].
\end{eqnarray}
Here ${\mathcal{H}}={\mathcal{H}}_0 + U({\mathbf{r}})$ and the matrix
elements are defined as $\langle nl|\hat{A}|n'l
\rangle\equiv\int_0^\infty
{\mathcal{R}}_{nl}(r)\hat{A}{\mathcal{R}}_{n'l}(r) r^2dr$.

We will examine certain regimes for which one can derive approximate
analytical results from Eq.(\ref{BdGeqnexpand}) [or Eq.(\ref{BdGeqn})]
and thus for the Cooper pairing in a harmonic trap. We keep the
temperature $T=0$ for the remaining part of this paper in order to
examine the ground state properties of the gas.  The nature of the
paired state depends on the relative magnitude of the Hartree field
$U(r)$, the trap level spacing $\hbar\omega$, and the pairing field
$\Delta(r)$.

\section{Intra shell pairing}\label{intrashellsection}
In this section, we examine the case when the interaction is so weak, that it only 
induces pairing between particles within the same harmonic oscillator shell [$n=n'$ 
in Eq.(\ref{BdGeqnexpand})], hence the term \emph{intra} shell pairing. In this regime
the pairing energy is smaller than the trap level spacing $2\hbar\omega$ between 
states with the same 
angular momentum quantum number $l$ [The Cooper pairing is between states with the 
angular quantum numbers $(l,m)$ and $(l,-m)$].
We ignore for the time being the splitting of the 
harmonic oscillator shells due to the Hartree field. The harmonic oscillator shells
therefore have the energies $\tilde{\xi}_{nl}\simeq\xi_n=(n+3/2)\hbar\omega-\mu_F$  with the angular 
momentum quantum number $l=n, n-2\ldots ,0$ ($n$ even) or $l=n, n-2 \ldots ,1$ ($n$ odd).
 The strength of the pairing between states 
with quantum numbers $(n,l)$ and $(n',l)$ in Eq.(\ref{BdGeqnexpand}) is given by 
the matrix element 
\begin{equation}\label{matrixelement}
\langle nl|\Delta(r)|n'l \rangle=\int_0^\infty dr r^2 \Delta(r){\mathcal{R}}_{nl}(r){\mathcal{R}}_{n'l}(r).
\end{equation}
The radial functions ${\mathcal{R}}_{nl}(r)$ in general change sign as a function of $r$.
In the ground state however, $\Delta(r)$ has a definite sign (chosen to be positive in the present case) as 
any oscillations in the pairing field in general costs energy. We therefore see that the ``inter-shell'' matrix 
elements with $n\neq n'$ will be suppressed due to the oscillating phase of ${\mathcal{R}}_{nl}(r){\mathcal{R}}_{n'l}(r)$ 
as compared to the ``intra-shell'' matrix elements with $n=n'$. The weak pairing regime considered 
in this section is characterized by the fact that the inter-shell matrix elements can be ignored. 
A simple perturbative calculation reveals that the parameter determining the importance 
of the inter-shell matrix elements is $|\langle nl|\Delta(r)|n'l \rangle|/\hbar\omega$. When this 
parameter is small these matrix elements can be ignored and we are in the intra-shell pairing regime. 
For stronger coupling, this parameter is no longer small and the inter-shell matrix elements changes the 
solution to the self-consistent and thus non-linear gap equations qualitatively. We will describe this in detail 
in sec.\ \ref{Strongpairing}.

Ignoring all inter-shell 
matrix elements, Eq.(\ref{BdGeqnexpand}) splits into  simple $2\times2$ matrix equations 
for each pair of quantum numbers $(n,l)$. For a given set of quantum numbers $(n,l)$, we obtain 
the usual solution $u_{nl}=(1+\xi_n/E_{nl})/2$, $v_{nl}=(1-\xi_n/E_{nl})/2$ with 
 $E_{nl}=(\xi_n^2+\Delta_{nl}^2)^{1/2}$ and 
 $\Delta_{nl}=\langle nl|\Delta(r)|nl \rangle$. Ignoring for the moment the regularization 
of the gap function, the self-consistent equation for the pairing field becomes:
\begin{equation}\label{gapequation}
\Delta(r)=\frac{|g|}{2}\sum_{n,l}\frac{2l+1}{4\pi}\frac{\Delta_{nl}}{E_{nl}}{\mathcal{R}}_{nl}^2(r)
\end{equation}
where the factor $2l+1$ comes from the summation over the angular quantum number $m$.
The nature of the self-consistent solution of the BdG equations depends on the position 
of the chemical potential $\mu_F$ relative to the trap levels. 
We will now examine two opposite cases of interest.

\subsection{Partly  filled Shell}\label{partlyfilled}
When the chemical potential is at a harmonic oscillator shell, i.e.\ 
 $\mu_F=(n_F+3/2)\hbar\omega$, this shell is partly filled for $T=0$. 
This is the typical situation physically, whereas the case of a completely filled shell 
corresponding to $\mu_F=(n_F+2)\hbar\omega$ at $T=0$ occur only for ``magic''
numbers of trapped particles.
In this section we will provide a analytical solution to the pairing problem when the interaction 
is so weak that only intra shell pairing is significant and  $\mu_F=(n_F+3/2)\hbar\omega$.
We will show that for very weak interaction the main contribution to the pairing correlations originate from the 
partly  filled shell at the chemical potential ($n=n_F$), whereas for stronger interactions the contribution
to the pairing from shells away from the chemical potential ($n\neq n_F$) is dominant.
Using Eq.(\ref{matrixelement}) and Eq.(\ref{gapequation}), the self-consistent equation for the intra shell pairing 
matrix elements  reads
\begin{gather}\label{pairingmatrixelement}
\Delta_{nl}=\frac{|g|}{2}\sum_{l'}\frac{2l'+1}{4\pi}\int_0^\infty dr r^2 {\mathcal{R}}_{nl}^2(r){\mathcal{R}}_{n_Fl'}^2(r)+
\nonumber\\
\frac{|g|}{2}\sum_{n'\neq n_F,l'}\frac{2l'+1}{4\pi}
\frac{\Delta_{n'l'}}{\sqrt{{\xi_{n'}}^2+\Delta_{n'l'}^2}}\int_0^\infty dr r^2
 {\mathcal{R}}_{nl}^2(r){\mathcal{R}}_{n'l'}^2(r)
\end{gather}
where the first term  singles out the contribution from the shell at the chemical potential. 
Notice  that for this shell,  $\xi_{n_Fl}=0$ such that $E_{n_F}=\Delta_{n_Fl}$.
Equation (\ref{pairingmatrixelement})
gives the strength of the pairing as a function of the angular momentum $l$ of the particles 
forming the Cooper pair and their shell quantum number $n$.
The $l$-dependence of the pairing, determined through the integral $\int dr r^2 \Delta(r){\mathcal{R}}_{nl}^2(r)$, 
is rather weak; we  will later show that 
$|\Delta_{n,l=n}-\Delta_{n,l=0}|/\Delta_{n,l=0}\ll 1$.
It is therefore not important for the present analysis and we 
define the $l$-averaged pairing strength as $\Delta_n=\sum_l(2l+1)\Delta_{nl}/\Omega_n$ 
with $\Omega_n=\sum_l(2l+1)=(n+1)(n+2)/2$  being the (single spin) degeneracy of the 
harmonic oscillator shell with energy $(n+3/2)\hbar\omega$.  For $n\gg 1$, we can 
use the Thomas-Fermi identity 
\begin{eqnarray} \label{Trick}
   \sum_l\frac{(2l+1)}{4\pi}{\mathcal{R}}_{n'l}^2(r) =
   \frac{\partial\rho_\sigma(r)}{\partial n'}=
   \frac{\sqrt{2n'+3}}{2\pi^2a_{osc}^3}\sqrt{1-r^2/R^2_{TF}(n')}
\end{eqnarray}
with $R_{TF}(n')=\sqrt{n'+3/2}a_{osc}$.
Here, $\rho_\sigma$ is the density from a single hyperfine state when shells up to energy
 $(n'+3/2)\hbar\omega$  are occupied;  $\partial\rho_\sigma(r)/\partial n'$ 
therefore should be understood as the derivative of the density 
with respect to the highest harmonic oscillator 
shell occupied.
The pairing strength $\Delta_n$ depends weakly on $n$ for the shells around $n=n_F$ which 
mainly contribute to the pairing and we therefore approximate 
$\Delta_n\simeq\Delta_{n_F}\equiv\Delta$.
Equation (\ref{pairingmatrixelement}) then becomes 
\begin{equation}\label{intragapeq}
   \Delta=G+\nu G
   \sum_{n\neq n_F}^{n<2n_F}\frac{\Delta}
  {\sqrt{[(n-n_F)\hbar\omega]^2+\Delta^2}} \,,
\end{equation}
where we have defined 
\begin{equation} \label{G}
   G=\frac{|g|}{2}\frac{\int dr r^2[\partial_n\rho_\sigma(r)|_{n_F}]^2}
   {\int dr r^2\partial_n\rho_\sigma(r)|_{n_F}}=
    \frac{32\sqrt{2n_F+3}}{15\pi^2} \frac{|a|}{a_{osc}}\hbar\omega \,.
\end{equation}
The correction factor $\nu$ arises because the overlap integrals in the
second term in Eq.(\ref{pairingmatrixelement}) depend on the shell
number $n'$. We approximate this dependence by averaging the factor
$\sqrt{n'}$, which gives
$\nu=\int_0^{2n_F}n^{1/2}dn/2n_F^{3/2}=2^{3/2}/3$.  We have introduced
the cut-off $n<2n_F$ in the sums which models as a first approximation the more rigorous regularization procedure
described in Ref.\ \cite{BruunBCS} that is required for a delta-function pseudo-potential. Such a procedure
can for the present purpose be approximated by a cut-off of $n<2n_F$ essentially due to  the fact that 
the Lippman-Schwinger eqn.\ for  two particle scattering appearing in the regularization procedure contains a single 
particle Green's function which is odd around the Fermi surface 
whereas the contribution to the pairing is even due to the particle-hole symmetry. 

When $\Delta\ll \hbar\omega$ we can expand and find the gap
\begin{equation}\label{finalgap1}
    \Delta=\frac{G}{1-2\nu
           \ln(e^\gamma n_F)\frac{G}{\hbar\omega}} \,.
\end{equation}
Here, $\gamma\simeq0.577$ is Eulers constant.
For very weak interactions, $G\ln(n_F)\ll\hbar\omega$, the gap is simply
$\Delta=G$. This corresponds to the limit where pairing only occurs in
the shell right at the chemical potential ($n=n_F$). The result
$\Delta=G$ was also obtained in ref.\cite{Heiselberg} using a
different approach (the seniority scheme). 
As $G\ln(n_F)$ increases the gap increases and $\Delta>G$ due to
intra shell pairing in shells away from the chemical potential.
 For $G\ln(n_F)\sim 1/2$, Eq. (\ref{finalgap1}) diverges.
However, at the same time the shell pairing ansatz  leading to Eq. (\ref{pairingmatrixelement}) breaks down
as will be described in detail in the following section.

\begin{figure}
\centering
\epsfig{file=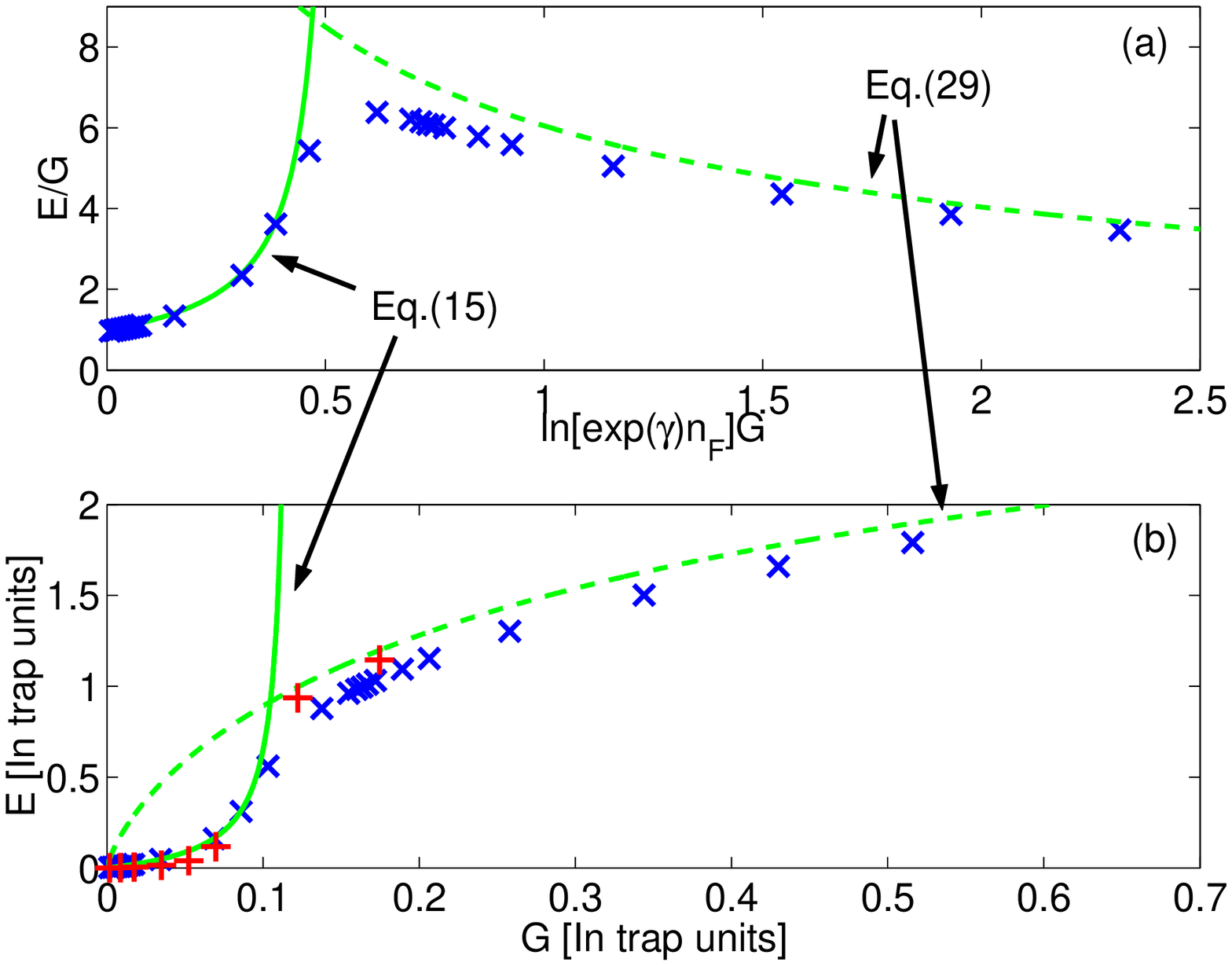,width=0.5\textwidth,height=0.6\textwidth}
\caption{The lowest quasi particle energy $E_{l=0}$ 
divided by the effective coupling strength $G$
as a function of $G\ln[\exp(\gamma)n_F]$ (in units of $\hbar\omega$) (a). Figure (b) depicts 
$E/\hbar\omega$ as a function of $G/\hbar\omega$.
 $\mu_F=51.5\hbar\omega$. Plusses and crosses are with and without 
the  Hartree field respectively.}
\label{mu51.5Hogaasen}
\end{figure}
In fig.\ (\ref{mu51.5Hogaasen}a), we plot the lowest $l$ averaged
quasi particle energy $E=\Delta$ (since $\xi_{n_F}=0$) divided by the
effective coupling strength $G$ as a function of
$G\ln[\exp(\gamma)n_F]$.  The solid curve is calculated from
Eq.(\ref{finalgap1}) whereas the $\times$'s are results obtained from
numerical calculations as described in ref.\ \cite{BruunBCS}.  We have
chosen $n_F=50$ for the numerical calculations corresponding to
$\sim2\times10^4$ particles trapped in each of the two hyperfine states. Note
that according to Eq.(\ref{finalgap1}), $E/G$ in terms of
$G\ln[\exp(\gamma)n_F]$ is a \emph{universal} function
independent of the size of the system (i.e.\ independent of $n_F$) as
long as the system is in the intra shell pairing regime. In this
regime, we see that there is excellent agreement between
Eq.(\ref{finalgap1}) and the numerical results. For stronger coupling,
the intra shell pairing assumption breaks down and the numerical
results differ qualitatively from Eq.(\ref{finalgap1}). The nature of
the Cooper pairs undergoes a transition to a region where the pairing
between particles in different harmonic shells is significant. The
transition in the nature of the pairing is clearly visible in fig.\
(\ref{mu51.5Hogaasen}a) since $E/G$ changes from being an increasing
to a decreasing function of $G\ln[\exp(\gamma)n_F]$ when the
intershell pairing becomes dominant.  We will examine the intershell
pairing in detail in section \ref{Strongpairing}. Of course, the
pairing strength is a monotonic increasing function of the coupling
strength $G$ as shown in fig.\ (\ref{mu51.5Hogaasen}b).
For weak pairing, [$G\ln(n_F)\ll\hbar\omega$] we find $\Delta= G$ in
agreement with the analysis above.  In this
regime there is only significant pairing in the shell at the chemical
potential. For stronger interaction with $G\ln(n_F)\gtrsim0.1\hbar\omega$, the Cooper pairing is still only
between particles within the same shell, but shells away from the
chemical potential ($n\neq n_F$) contribute significantly to the
pairing energy as predicted by Eq.(\ref{finalgap1}).

The reason that  shells away from the chemical potential contribute to the pairing for stronger coupling 
is that it becomes energetically favorable to partially occupy these shells in order to 
make a coherent state with stronger spatial overlap between the particles forming the Cooper pairs.

We will now briefly examine the dependence of the pairing strength $\Delta_{n_Fl}$ given 
by Eq.(\ref{pairingmatrixelement}) on the angular momentum quantum 
number $l$. It is assumed that we are in the very weak coupling regime [$\ln(n_F)G\ll 1$] such that 
 such that only the shell at the chemical potential $n=n_F$ contributes to the pairing. 
 Using  Eq.(\ref{Trick}), 
${\mathcal{R}}_{n_Fl=n_F}(r)\propto r^{n_F}\exp(-r^2/2a_{osc}^2)$, 
and the semiclassical approximation
$r{\mathcal{R}}_{n_Fl=0}(r)\simeq [k(r)]^{-1/2}\sin(\int^rdr'k(r')+\phi)$ with 
$k(r)=\sqrt{2n_F-r^2/a_{osc}^2}$, the radial 
integral in Eq.(\ref{pairingmatrixelement}) can easily be solved for $l=0$ ($n_F$ even) and $l=n_F$ yielding
\begin{equation} \label{ldependence}
\frac{E_{n_Fl}}{G}=\frac{\Delta_{n_Fl}}{G}=
\left\{\begin{array}{ccl}
15/16&,&l=0\\
15\pi/(32\sqrt{2})&,& l=n_F
\end{array}
\right.
\end{equation}
with $G$ given by Eq.(\ref{G}). From Eq.(\ref{ldependence}), we see that 
$|\Delta_{n_F,l=n_F}-\Delta_{n_F,l=0}|/\Delta\simeq0.10\ll 1$ thereby justifying the
 $l$ averaging of the pairing strength described above. 

\begin{figure}
\centering
\epsfig{file=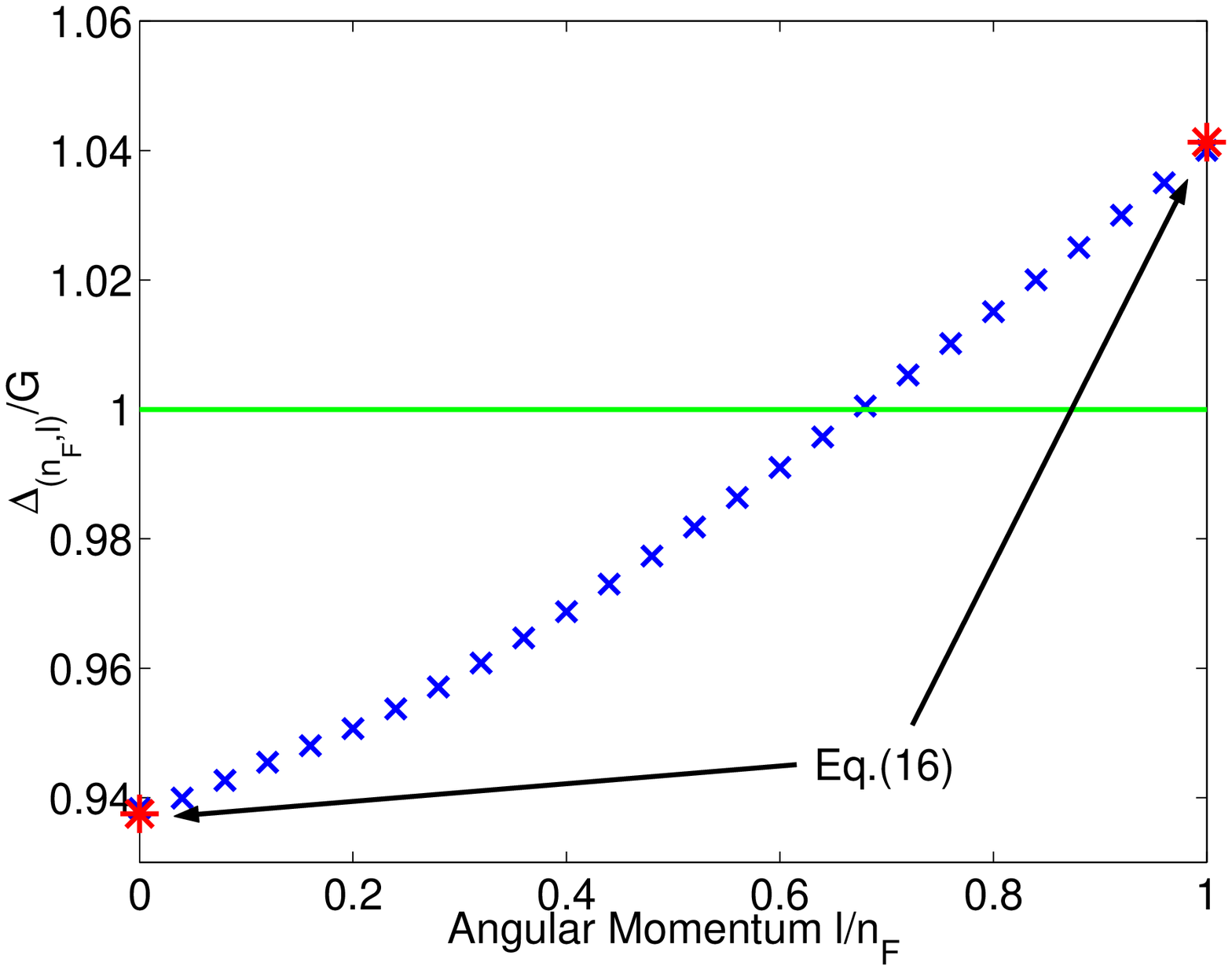,height=0.25\textwidth,width=0.5\textwidth}
\caption{The lowest quasi particle energies  $E_{n_Fl}=\Delta_{n_Fl}$ (in units of $G$) as a function of 
the angular momentum $l/n_F$ for $\mu_F=51.5\hbar\omega$ and $G=1.75\times10^{-4}\hbar\omega$.}
\label{ldependencefig}
\end{figure}
Figure (\ref{ldependencefig}) shows the $l$-dependence of the lowest
quasi particle excitation energy $E_{n_Fl}=\Delta_{n_Fl}$ for the same
parameters as for fig.\ (\ref{mu51.5Hogaasen}) and
$G/\hbar\omega=1.75\times10^{-4}\ll\ln(n_F)^{-1}$.  The $\times$'s
indicate the numerical results, the $*$'s indicate the results given
by Eq.(\ref{ldependence}). We also plot the $l$ averaged value
$\Delta=G$ as a solid line. The numerical results agree very well with the analytical
predictions of Eq. (\ref{ldependence}) up to $1/n_F$ corrections.
We notice that the $l$-dependence is weak thereby justifying the $l$-averaging of the
pairing assumed above. Again, one should note that fig.\
(\ref{ldependencefig}) is universal in the sense that it is
independent of the size of the system given by $n_F$; one always has
$0\le l/n_F\le1$ and $E_{n_Fl}/G$ is independent of $n_F$ for
$\ln(n_F)G\ll 1$ as can be seen from Eq.(\ref{ldependence}).

\subsection{Filled Shell}\label{filled}
We now examine the case when the chemical potential is exactly between
two harmonic oscillator shells, i.e.\ $\mu_F=(n_F+2)\hbar\omega$. This
case corresponds to shells up to and including $n=n_F$ being
completely filled and shells with $n>n_F$ empty when there is no
pairing. As we will see, this situation differs qualitatively from the
case considered in sec.\ (\ref{partlyfilled}) in the sense that the
pairing is zero below a certain critical coupling strength
$G_c$. Again, we are in this section in the weak pairing regime 
such that only intra shell pairing is significant. Thus,
Eq.(\ref{gapequation}) is still valid and leads
to a gap equation analogous 
to Eq.(\ref{pairingmatrixelement}), however, excluding the first term
 corresponding to pairing in the shell at the
chemical potential. It is the absence of this term that
leads to the vanishing of pairing below a certain finite critical
coupling strength. Following steps completely analogous to the ones
leading from Eq.(\ref{pairingmatrixelement}) to Eq.(\ref{intragapeq}),
we obtain the gap equation
\begin{equation}\label{intragapeq2}
   \frac{\hbar\omega}{\nu G} = 
   \sum_{n=0}^{2n_F}\frac{1}{\sqrt{(n-n_F-1/2)^2+(\Delta/\hbar\omega)^2}}
\end{equation}
with $G$ still defined by Eq.(\ref{G}) and the scaling factor $\nu=2^{3/2}/3$ again approximating the 
$n$ dependence of the 
radial overlap integrals. The critical coupling strength $G_c$ below which 
there is no pairing is found by setting $\Delta=0$ in Eq.(\ref{intragapeq2}). Neglecting 
terms of ${\mathcal{O}}(n_F^{-1})$, we obtain
\begin{equation}\label{Gcritical}
   G_c=\frac{3\hbar\omega}{2^{5/2}\ln(4e^\gamma n_F)}.
\end{equation}
Using Eq.(\ref{Gcritical}) and expanding Eq.(\ref{intragapeq2}) in the small parameter 
$\Delta/\hbar\omega$, we find
\begin{equation}\label{finalgap2}
   \frac{\Delta}{\hbar\omega}=
   \left\{\begin{array}{lcl}
   0&,&G<G_c\\
   \left[\frac{3\hbar\omega}{2^{3/2}7\xi(3)}\right]^{1/2}
   \sqrt{\frac{1}{G_c}-\frac{1}{G}}&,& G\ge G_c 
\end{array}
\right.
\end{equation}
with $\xi(3)=\sum_{n=1}^\infty n^{-3}\simeq1.202$. Equation (\ref{finalgap2}) should be compared
to Eq.(\ref{finalgap1}). We see that the main difference is that when
there is a harmonic oscillator shell at the chemical potential, the
pairing occurs as soon as there is an infinitesimal attraction between
the particles. When the chemical potential is between two shells,
there is only pairing when the coupling is above a certain critical
coupling strength $G_c$ as there is zero density of states at the
chemical potential. 

\begin{figure}
\centering
\epsfig{file=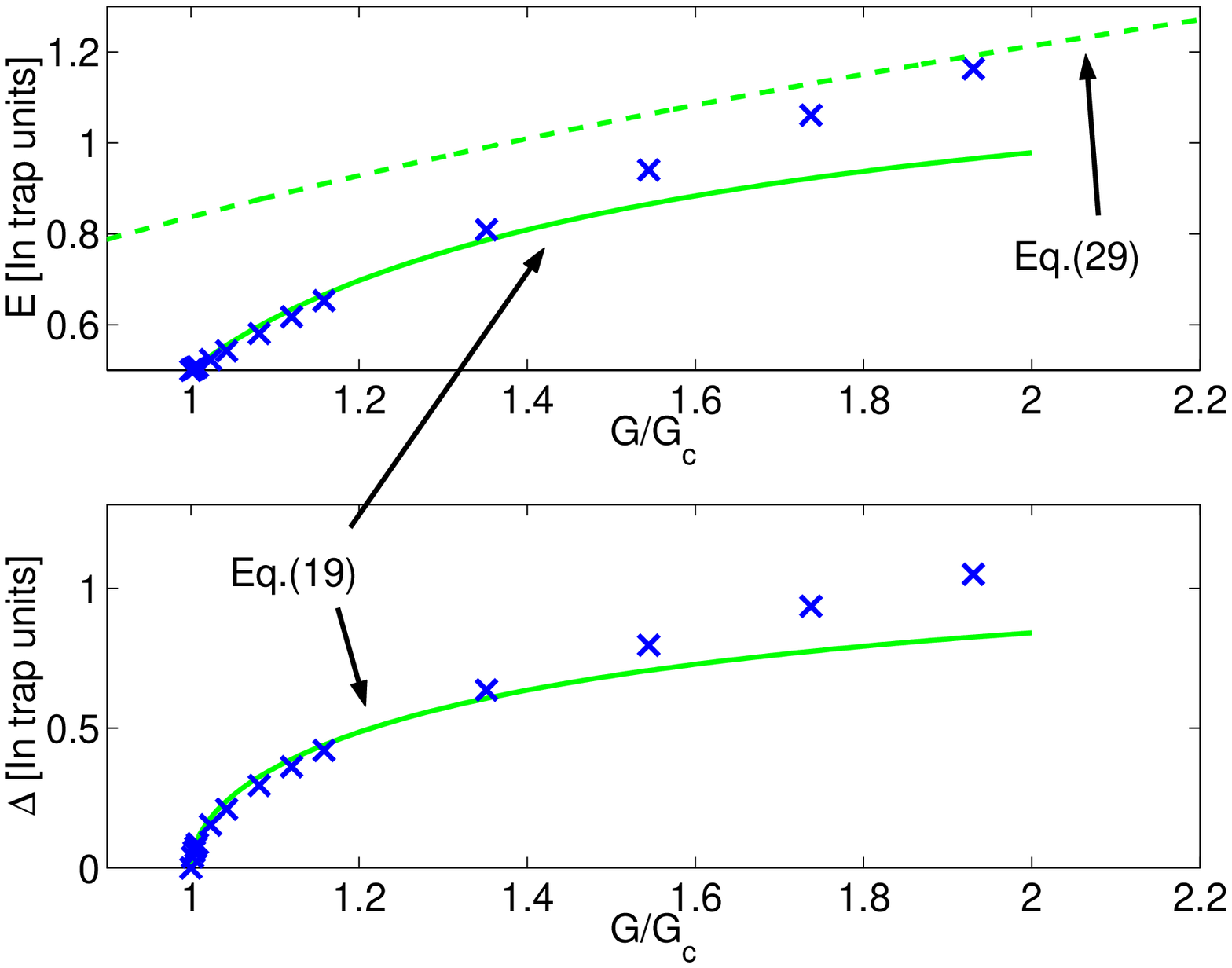,height=0.4\textwidth,width=0.5\textwidth}
\caption{The lowest $l$ averaged quasi 
particle energy $E/\hbar\omega$ (a) and the pairing 
strength  $\Delta/\hbar\omega=[E^2-(\hbar\omega)^2/4]^{1/2}/\hbar\omega$ as a function of $G/G_c$.}
\label{mu52Hogaasen}
\end{figure}
In analogy with fig.(\ref{mu51.5Hogaasen}), we plot in fig.(\ref{mu52Hogaasen}) the lowest $l$ averaged quasi 
particle energy as a function of $G/G_c$. The line is Eq.(\ref{finalgap2}) and  the 
$\times$'s are obtained from numerical calculations with $\mu_F=52\hbar\omega$. As before, we conclude 
from Eq.(\ref{finalgap2}) that $\Delta$ as a function of $G/G_c$ is independent of $n_F$ in the intra-shell 
pairing regime. Of course, the size of the system is important when determining $G_c$ as is apparent from 
Eq.(\ref{Gcritical}).
Note that the lowest quasi particle energy $E$ is calculated as 
$E=(\xi_{n_F}^2+\Delta^2)^{1/2}$ with $\xi_{n_F}=\hbar\omega/2$ and $\Delta$ given 
by  Eq.(\ref{finalgap2}). To enable a more direct comparison between the numerical 
results for the pairing strength and Eq.(\ref{finalgap2}), we also plot 
 $\Delta=(E^2-\xi_{n_F}^2)^{1/2}$. The observable in an experiment that probes single
particle excitations is of course the 
excitation energy $E$ and not $\Delta$. We see that there is excellent agreement between the 
numerical results and  Eq.(\ref{Gcritical})-(\ref{finalgap2}) for $\Delta/\hbar\omega\lesssim 1/2$. The
 approximate analytical 
formulae predict very well the critical coupling strength and the size of the pairing 
for small $\Delta/\hbar\omega$ where only intra shell pairing is significant. 
As in sec.(\ref{partlyfilled}), the pairing between 
particles in different harmonic oscillator shells become important when 
 $\Delta/\hbar\omega\sim {\mathcal{O}}(1)$ and the intra shell pairing theory breaks 
down. 

Pairing between just two shells was studied in the context of nuclear
structure where similar results were found.
Comparing to experimental data~\cite{BM} only qualitative agreement
was found. 
The nuclear spectra are complicated by stronger anharmonic mean field potentials and the finite
range of the nucleon-nucleon interaction as well as  spin and spin-orbit dependences.
 For atomic traps the results presented here should be more
accurate as the pseudo potential approximation for the atom-atom
interaction is expected to work very well for low atomic densities,
and since the mean field is weaker.

\subsection{Effect of the Hartree field}
We have sections \ref{partlyfilled}-\ref{filled} ignored the  Hartree field which induces 
an energy shift $\tilde{\xi}_{nl}$ as given by Eq.(\ref{Hartree}) 
of the levels within a given harmonic oscillator shell. In this section, we will examine 
the effect of this mean field induced spread in the harmonic oscillator shell energies 
on the pairing.
The gap equation including the Hartree field becomes in complete analogy with Eq.(\ref{pairingmatrixelement})
\begin{equation}\label{pairHartree}
\Delta_{nl}=\frac{|g|}{2}\sum_{n',l'}\frac{2l'+1}{4\pi}
\frac{\Delta_{n'l'}}{E_{n'l'}}\int_0^\infty dr r^2
\tilde{R}_{nl}^2(r)\tilde{R}_{n'l'}^2(r)
\end{equation}
with quasi particle energies
\begin{equation}
 E_{nl}= \sqrt{\tilde{\xi}_{nl}^2+\Delta_{nl}^2}
\end{equation}
and $\tilde{R}_{nl}(r)$ being the solution to the radial Schr\"{o}dinger 
equation including the Hartree field. 
 From Eq.(\ref{Hartree}), we see that the energy spread of a given 
shell due to the Hartree field is 
 $|\tilde{\xi}_{nl=0,1}-\tilde{\xi}_{nl=n}|\propto n^{3/2}a/a_{osc}$ whereas the size of the pairing 
assuming completely flat shells is given by Eq.(\ref{finalgap1}) [or Eq.(\ref{finalgap2})]. Thus, for large $n_F$ 
the Hartree field splits the different $l$ levels within the same shell
sufficiently so that the pairing occurs only within 
one (or a few) energy levels $\tilde{\xi}_{n_Fl}$ close to the chemical potential~\cite{Heiselberg}. 
Assuming that one level $\tilde{\xi}_{n_Fl}$ is at the chemical potential (i.e.\ $\tilde{\xi}_{n_Fl}=0$),
only this level will contribute to the pairing for weak coupling and  Eq.(\ref{pairHartree}) becomes
\begin{equation}
\Delta_{n_Fl}=\frac{|g|}{2}\frac{2l+1}{4\pi}
\int_0^\infty dr r^2\tilde{R}_{n_Fl}^4(r).
\end{equation}
This equation is straightforward to solve as we simply have to evaluate 
the radial integral. For instance, for $l=n_F$, we obtain:
\begin{equation}\label{singlelgap}
\frac{\Delta_{n_Fl=n_F}}{\hbar\omega}=
\frac{|a|/a_{osc}}{\sqrt{\pi}}.
\end{equation}
Equation (\ref{singlelgap})  agrees with the results derived in ref.\ \cite{Heiselberg} using 
the seniority scheme. 

For Eq.(\ref{singlelgap}) to be valid, we need $\Delta_{n_Fl}\ll
|\tilde{\xi}_{n_Fl}-\tilde{\xi}_{n_Fl\pm2}|$. Otherwise, the levels
$\tilde{\xi}_{n_Fl\pm2\pm4\ldots}$ away from the chemical potential
will contribute significantly to the pairing and the assumption that
only particles in the level $\tilde{\xi}_{n_Fl}=0$ form Cooper pairs leading to
Eq.(\ref{singlelgap}), will be incorrect. In the discussion following
Eq.(\ref{finalgap1}), we concluded that for
$\ln(n_F)G\gtrsim0.1\hbar\omega$ levels away from the chemical
potential with normal phase energies $\xi_n=(n-n_F)\hbar\omega$
contributed significantly to the pairing (see Fig.\
\ref{mu51.5Hogaasen}). Equation (\ref{pairHartree}) with one level at
the chemical potential ($\tilde{\xi}_{n_Fl}=0$) is completely
analogous to Eq.(\ref{pairingmatrixelement}); ignoring the
contribution to the pairing from $n\neq n_F$ shells in
Eq.(\ref{pairHartree}) the essential difference between the two
equations is the level spacing between the normal phase energies which
is $\hbar\omega$ in Eq.(\ref{pairingmatrixelement}) and
$\tilde{\xi}_{n_Fl}-\tilde{\xi}_{n_Fl\pm2}$ in
Eq.(\ref{pairHartree}). Thus, a similar analysis on
Eq.(\ref{pairHartree}) yields that the contribution from states
$\xi_{n_Fl'}\neq 0$ to the pairing can be ignored if
$|\tilde{\xi}_{n_Fl}-\tilde{\xi}_{n_Fl\pm2}|\gg \Delta_{n_Fl}\ln(n_F)$
with $\Delta_{n_Fl}$ given by Eq.(\ref{singlelgap}). For $l=n_F$ which
gives the maximum $|\tilde{\xi}_{n_Fl}-\tilde{\xi}_{n_Fl\pm2}|$, this
condition yields [using Eq.(\ref{Hartree}) and Eq.(\ref{singlelgap})]
$n_F\gg 200$.  That is, only for rather large systems with $\mu_F\gg
200\hbar\omega$ (corresponding to $N\gg10^6$ particles trapped), is
the pairing for weak coupling only significant in the level
$\tilde{\xi}_{n_Fl_F}=0$ at the chemical potential and
Eq.(\ref{singlelgap}) is valid. If the chemical potential is at a
level with $l<n_F$ (i.e.\ $\tilde{\xi}_{n_Fl}=0$ with $l<n_F$), the
Hartree splitting of the levels is smaller than for $l=n_F$ and the
condition for single $(n_F,l)$ pairing becomes even more
strict~\cite{Heiselberg}.

The assumption in sec.\ref{partlyfilled}-\ref{filled} that the Hartree
field can be neglected such that the harmonic oscillator shells are
completely flat, is consistent if
$\Delta>|\tilde{\xi}_{n_Fl=0,1}-\tilde{\xi}_{n_Fl=n_F}|$ with $\Delta$
given by Eq.(\ref{finalgap1}) [or Eq.(\ref{finalgap2})]. From
Eq.(\ref{Hartree}), we have
$|\tilde{\xi}_{n_Fl=0,1}-\tilde{\xi}_{n_Fl=n_F}|\simeq(5/32)n_FG$.
Using this and Eq.(\ref{finalgap1}) in the very weak coupling regime
with $\Delta=G$, we get the condition $n_F\lesssim 6$ for the validity
of neglecting the Hartree field. For stronger coupling where pairing
in the shells away from the chemical potential is important, we of
course obtain that the neglect of the Hartree field is consistent for
larger $n_F$, since Eq.(\ref{finalgap1}) gives $\Delta>G$.  In
particular, for $\Delta\sim0.5\hbar\omega$ which is the approximate
maximum value for which the intra shell pairing assumption holds, we
obtain that Eq.(\ref{finalgap1}) is valid for systems with
$n_F\lesssim 30$.

In order to examine the influence of the Hartree field, we have plotted
in fig.(\ref{mu51.5Hogaasen})(b) as $+$'s the lowest quasi particle energy
(for which $\tilde{\xi}_{nl}=0$ such that $E_{n_Fl}=\sqrt{\Delta_{n_Fl}^2+\tilde{\xi}_{nl}^2}=\Delta_{n_Fl}$)
including the effect of the Hartree field in order to compare to the
results neglecting the Hartree field.  We keep the number of particles
trapped the same with or without the Hartree field.  Since
$n_F=50>30$, according to the analysis above the inclusion of the Hartree
field should yield a significant modification of the pairing as
compared to the analysis presented in sec.\ref{partlyfilled}.  As
expected, we see that in the intra shell pairing regime the Hartree
field suppresses the pairing as it introduces a splitting of the
levels in the shell at the chemical potential such that the density of
states for $\tilde{\xi}_{nl}=0$ is reduced. 
 For stronger interaction, where there is pairing between particles in different
harmonic oscillator shells, the effect of the Hartree field is to
increase the pairing strength. This is because in this regime the
Hartree field compresses the gas increasing the density $\rho(r=0)$
thereby increasing the effective interaction strength $k_F(r=0)|a|$
[See Eq.(\ref{rho})].  We see that Eq.(\ref{finalgap1}) gives a
qualitatively correct description of the pairing strength for this set
of parameters since $n_F=50$ is an intermediate value where the
pairing in the intra shell regime is not in the limit where the
Hartree field can be ignored, but also far away from the limit where
there is only pairing within the level $\tilde{\xi}_{nl}=0$.

\section{Pairing in the $\Delta(r=0)\gg\hbar\omega$ limit}\label{Strongpairing}
In this section, we will derive an analytical expression for the
lowest single particle energy in the regime, where the pairing field
$\Delta(r)$ is much larger than $\hbar\omega$, yet still in the
dilute limit so that $\Delta(r)\la E_F$. We will show that
the strong pairing between particles in different shells leads to a break-down of intra shell
pairing ansatz of secs.\ref{partlyfilled}-\ref{filled}; we
are in the \emph{inter} shell pairing regime.  

When $\Delta(r=0)\gg\hbar\omega$ the local or TF
coherence length in the center of the trap~\cite{deGennes} 
\bea
    \xi(r=0)=\frac{k_F(r=0)}{\pi m\Delta(r=0)} 
\eea
is smaller than the size of the system. The coherence length in general determines the 
length scale over which the pairing field can vary. In this regime, the shell
structure of the trap levels can be ignored and the gas behaves in
many ways as a quasi-homogeneous bulk system.  In nuclei gaps are
always smaller than $\hbar\omega$ by an order of magnitude \cite{BM}.
Thus trapped atoms provide a bridge between condensed matter and
nuclear physics for studying superfluid properties.

\begin{figure}
\centering
\epsfig{file=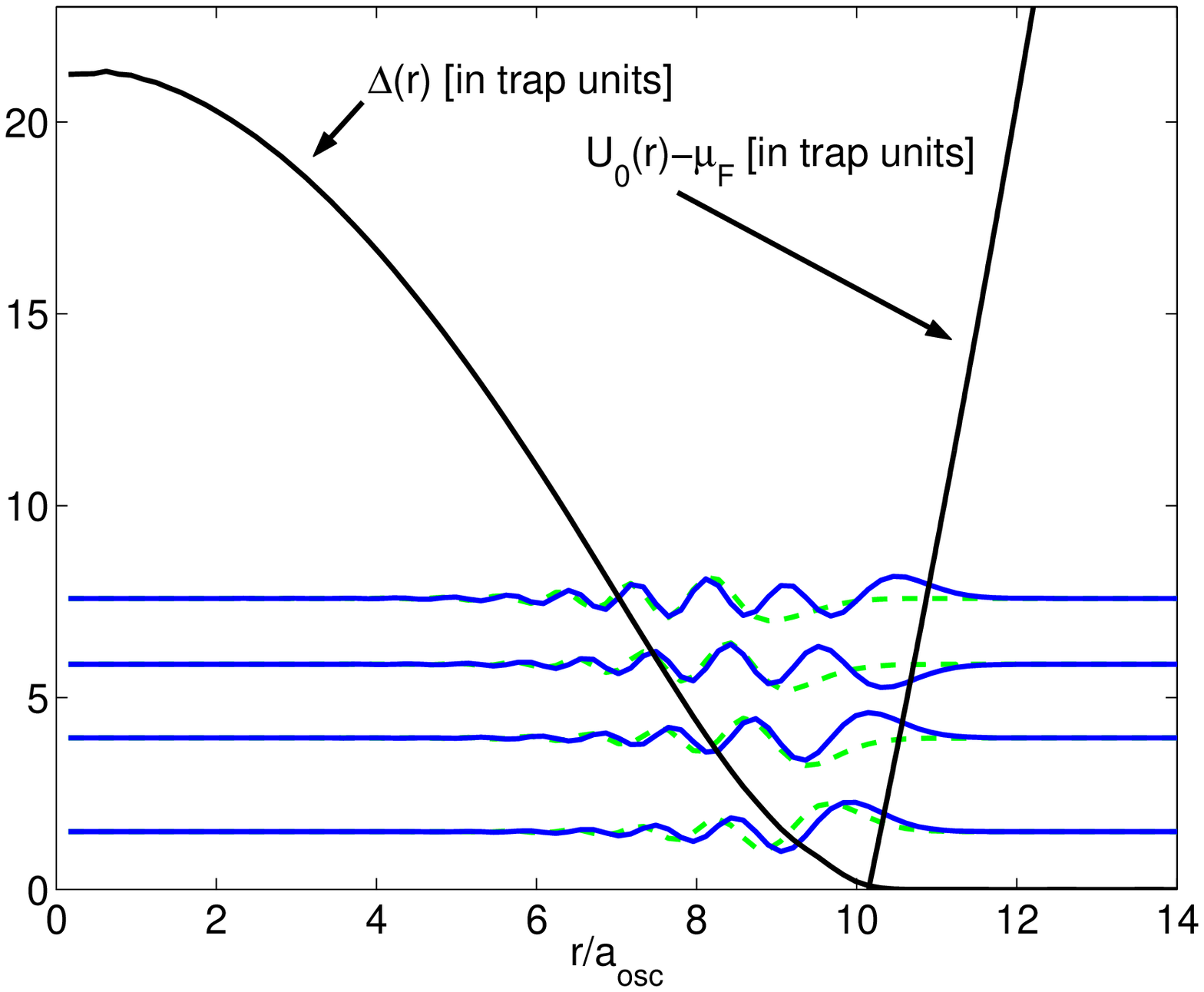,width=0.5\textwidth,height=0.35\textwidth}
\caption{The lowest 4 Bogoliubov wavefunctions $u_{\eta l}(r)$
 and $v_{\eta l}(r)$ in units of $a_{osc}^{-3}$ 
for $l=0$, $\Delta(r)$, and $U_0(r)-\mu_F$ in units of $\hbar\omega$.}
\label{UVplot}
\end{figure}

The pairing field $\Delta(r)$ in Eq.(\ref{BdGeqn}) in general increases with
$r\rightarrow 0$ in the bulk regime.  When $\Delta(r=0)\gg\hbar\omega$, we have
$\xi_{BCS}(r=0)/R_{TF}\ll1$. One can in this regime use a
semiclassical theory developed in Ref.\ ~\cite{Baranovsingle} to
obtain an analytical expression for the lowest quasi particle energy.
As discussed in Ref.\ ~\cite{Baranovsingle}, the lowest excitations
are concentrated in the region around the minimum of
$\Delta(r)+U_0(r)-\mu_F$. The lowest excitations have energies
$E_{\eta l}$ significantly lower than $\Delta(r=0)$ and can be denoted
``in gap excitations''.  Their wave-functions are suppressed in the
regions where $E_{\eta l}< \Delta(r)$ and $E_{\eta l}+\mu_F<
U_0(r)$. To illustrate this, we plot in Fig.\ (\ref{UVplot}) the
four Bogoliubov wave functions $[u_{\eta l}(r),v_{\eta
l}(r)]$ with lowest energy $E_{\eta l}$ 
for  $l=0$, $\mu_F=51.5\hbar\omega$ and $G\simeq0.34\hbar\omega$.
The solid lines are the $u_{\eta l=0}(r)$ and the dashed lines are the
$v_{\eta l=0}(r)$. The wave functions are centered horizontally along a $y$-value
equalling their excitation energy in units of $\hbar\omega$. We
also plot $\Delta(r)$ and $U_0(r)-\mu_F$ to illustrate how the lowest
quasi particle excitations with energies $E_\eta\ll\Delta(r=0)$ are
centered at the edge of the cloud where $\Delta(r)+U_0(r)-\mu_F$ has a
minimum. 

 For $E_\eta\ll\Delta(r=0)$, a semiclassical analysis yields the
following quantization condition for the $l=0$
states~\cite{Baranovsingle}:
\begin{equation}\label{semiclasquant}
 (\eta+\frac{1}{2}) \frac{\pi}{2}\hbar\omega =
 \int_{r_c}^{R_{TF}} dr\frac{\sqrt{E_\eta^2-\Delta(r)^2}}{\sqrt{R^2_{TF}-r^2}} 
\end{equation}
with $\eta=0,1,\ldots$. 
Here, $r_c$ is the left turning point determined by $\Delta(r_c)=E_\eta$. 
 For $\Delta(r=0)\gg\hbar\omega$ we have $1-r_c/R_{TF}\ll 1$. 
The centrifugal barrier, present for  $l>0$ plays a minor role
since the 
Bogoliubov wavefunctions are suppressed by $\Delta(r)$ in the region where the centrifugal 
potential is significant. 
Therefore the $l$-dependence of $E_\eta$ is weak and 
only when $l$ becomes comparable to $n_F$ does the centrifugal 
potential affect the lowest energy states for a given $l$ 
by increasing its energy. 

In a uniform system the pairing gap is 
\begin{equation}\label{DeltaBCS}
\Delta\simeq\kappa \frac{p_F^2}{2m}e^{-\pi/2k_F|a|} \,.
\end{equation}
 The prefactor $\kappa=8e^{-2}$ within the usual BCS theory but is
further reduced by a factor $(4e)^{-1/3}$, when induced interactions are
included~\cite{PethickBCS}.

In a non-uniform system such as the h.o. traps we can use the local
density (or TF) approximation 
in the $\Delta(r=0)\gg\hbar\omega$ regime, 
where the discrete level structure is of minor importance as compared to the
strong pairing field.
A Thomas-Fermi treatment yields 
\begin{equation}\label{Deltar}
\Delta(r)\simeq\kappa\mu_F(1-r^2/R_{TF}^2)e^{-\pi/2k_F(r)|a|} \,.
\end{equation}
The gap decreases rapidly near the surface and we can therefore neglect the
gap in the region $r_c<r<R_{TF}$ [see Fig.\ (\ref{UVplot})]. 
The semiclassical quantization condition thus simply gives
\begin{equation}\label{semiquantapp}
   E_{\eta } = \frac{\pi}{2} \frac{(\eta+\frac{1}{2})\hbar\omega}
                          {\arccos(r_c/R_{TF})}
\end{equation}
Here, the turning point given from
$E_\eta=\Delta(r_c)$ is now determined from the gap of Eq.(\ref{Deltar}).
We thus obtain in the $\Delta(r=0)\gg \hbar\omega$ regime
\begin{equation}\label{largeD}
   G \simeq \frac{32E_\eta}{15\pi^2(\eta+1/2)}
   \left\{\ln\left[\frac{\kappa(\eta+1/2)^2\pi^2}{4}\frac{n_F(\hbar\omega)^3}{E_\eta^3}\right]\right\}^{-1} \,,
\end{equation}
from which the quasi particle  particle energy $E_{\eta }$ can be
found by inversion from the coupling strength $G$ defined in Eq.(\ref{G}).

The lowest quasi particle energy, $E_0$, from Eq. (\ref{largeD}) (plotted as a dashed line)
is compared with the numerical results in
fig.(\ref{mu51.5Hogaasen}) and fig.(\ref{mu52Hogaasen}).  We see that
the agreement between the semi classical prediction and the numerical
results is good for $E\gtrsim \hbar\omega$ where $\Delta(r=0)\gg
\hbar\omega$.  Note that in the inter shell pairing regime, the
position of the chemical potential with respect to the Harmonic
oscillator levels is unimportant as the lowest quasi particle energy
is determined by the geometry of the minimum of
$\Delta(r)+U_0(r)$. Thus, Eq.(\ref{largeD}) agrees well with the
numerical results for the parameters relevant for both
fig.(\ref{mu51.5Hogaasen}) (partly filled shell) and fig.(\ref{mu52Hogaasen}) (filled shell).  Combined
with the theory presented in sec.(\ref{intrashellsection}), we
therefore have a theory describing the strength of the pairing fairly
well in the regimes $\Delta\lesssim\hbar\omega/2$ and
$\Delta(r=0)\gg\hbar\omega$. In the weak coupling regime, the pairing
is between particles within the same shell whereas for stronger
interaction, there is pairing between particles in different shells
and the lowest energy excitations are states localized in the surface
region around the minimum of $\Delta(r)+U_0(r)$.

Equation (\ref{largeD}) for the lowest quasi particle energy should be
compared with Eq.(\ref{Deltar}) giving the size of the pairing field
in the center of the trap. We see that contrary to the homogeneous
case, the lowest pairing energy in general is different from the size
of the pairing field even in the quasi homogeneous bulk limit; it grows
much slower with $G$ than $\Delta(r=0)$.

\section{Conclusion}\label{conclusion}
In this paper, we considered the properties of the
elementary excitations of a superfluid trapped atomic Fermi gas. The
nature of the paired state was shown to undergo transitions between
several different regimes as  the
number of particles (or the h.o. quantum number $n_F=(3N)^{1/3}$)
and the strength of the interaction are varied. Three regions emerge:

\begin{itemize}

\item
For weak coupling, $G\ln(n_F)\la 0.1$, the Cooper pairs were shown to be formed
between states in the h.o. shell at the Fermi surface
only. In this limit, it was
demonstrated that both the position of the chemical potential with
respect to the harmonic oscillator shells and the size of the Hartree
field, are strongly influencing the nature of the paired stated. 
The pairing gap is $\Delta=G$ for sufficiently few particles in the trap,
$n_F\la 10$, but for more particles the mean field splits
the single particle levels and reduce the pairing. For $n_F\gg 100$ the
pairing is reduced to that in a single $l$-level.

\item For interactions such that $0.1\la G\ln(n_F)/\hbar\omega\la 1/2$
Cooper pairs are still essentially only formed between states 
within the same shell. However, there is pairing in many shells besides
that at the Fermi level
yielding a gap significantly larger than $\Delta=G$.

\item For strong interactions, $G\ln(n_F)/\hbar\omega\ga 1/2$, 
the pairing field exceeds $\Delta(r=0)\ga\hbar\omega$.
The coherence length is shorter than the system size, and the gas 
can for many purposes be regarded as a quasi-homogeneous bulk system
The Cooper pairs are formed between
particles in different shells and the elementary excitations are
confined spatially to the surface of the gas.
\end{itemize}
It should be noted that the transitions between the regions described above
are smooth contrary to thermodynamic phase transitions. 
 The analytical formulae
appropriate for the various regimes were compared with numerical
calculations showing excellent agreement. 
 The results presented in this paper suggest, that by varying the number 
of particles and/or their interaction strength the trapped atomic gases
can behave either essentially as giant nuclei (Sec.\ref{intrashellsection})
 or as bulk systems (Sec.\ref{Strongpairing}). They thus provide an 
intriguing link between uniform systems explored in
condensed matter physics and finite systems relevant for nuclear
physics, quantum dots, etc.

\end{document}